\documentclass[aps,pra,twocolumn]{revtex4}
\usepackage{amsfonts}
\usepackage{bbding}
\usepackage{amssymb}
\usepackage{color}
\usepackage{amsmath}
\usepackage{graphicx}
\usepackage{subfigure}
\usepackage{txfonts}
\usepackage{bm}
\usepackage{ulem}
\usepackage[colorlinks,linkcolor=blue,citecolor=blue]{hyperref}
\newcommand{\ket}[1]{\left| #1 \right>}

\begin{document}
\title{Early-stage memory effect on the dephasing charger-mediated quantum battery}
\author{Yu Wang and Jiasen Jin}
\email{jsjin@dlut.edu.cn}
\affiliation{School of Physics, Dalian University of Technology, 116024 Dalian, China}
\date{\today}
\begin{abstract}
We investigate the performance of the charger-mediated quantum battery modeled by a two-qubit system. One of the qubits acts as the battery and the other acts as the charger which is subjected to a reservoir. We derived the time-local master equation in Lindblad form with a time-dependent dephasing rate. The dephasing rate may be negative in the early-stage of the charging process and thus indicate the presence of the memory effect. We find that such early-stage memory effect could increase the maximal ergotropy of the battery compared with the one under Markovian approximation with the corresponding asymptotic dephase rate. The enhancement of the performance is explained by means of the non-Markovian quantum jumps. Moreover, a discrete time scheme of the measurement-enhanced quantum battery is proposed in a quantum circuit with global and random local operations.
\end{abstract}
\maketitle
\section{Introduction}
\label{Introduction}

The quantum effects, including the quantum coherence and entanglement, become more and more pronounced in the micro- and nano scale systems \cite{horodecki2009rmp}. This has spurred the development of quantum devices which exhibit exotic properties and superior performance compared to their classical counterparts. Among the quantum devices, the quantum battery, as a device for temporarily storing and transferring energy, has attracted significant attention in the past years \cite{campaioli2024rmp}. The simplest model of a quantum battery in discrete-variable systems is the two-level system. A large variety of intriguing implementations of quantum battery with the charging and discharging process under the unitary dynamics have been reported in the single and many-body systems \cite{binder2015njp,ferraro2019prl,barra2019prl,rossini2020prl,francica2020prl,seah2021prl,chen2022pre,dou2022prb,dou2022pra,song2022pre,yan2023prappl,wang2024prl, liu2024pra,zhang2024pra,gyhm2024avsqs,wang2025prl,shastri2025npjqi,canzio2025pra,yao2025pra,hu2025,lu2025prl,yan2026prl,yan2026arxiv}.

The inevitable interaction between the quantum systems and its surrounding environments always leads to the loss of coherence, thus generally entails the degradation of the performance of a quantum battery \cite{breuer_book}. In contrast, the dissipation can also be viewed as a resource to boost the charging of quantum battery with a mediated system named as the charger \cite{farina2019prb}. Usually the dynamics of the charger and quantum battery can be described by the so-called Markovian process during which the information of the system flows to the environment. Indeed, the Markovian approximation is valid, at least for capturing the behavior of the long-time evolution when the time resolution is much larger than the time scale over which the environment correlation decays.

However, the non-Markovian dynamics is ubiquitous in practical quantum systems \cite{breuer2009prl,rivas2010prl}. The non-Markovianity can be used to enhance the performance of quantum batteries \cite{li2022,morrone2023qst,zhao2025pre,xu2026pra,li2026pre}. A typical example is the time-evolution of a qubit interacting with a structured environment which can be described by a Markovian master equation over a large time scale of evolution \cite{haikka2010,haikka2010physcri}. The non-Markovianity featured by the information backflow, although lasts shortly, is visible in the early stage of the evolution. For quantum batteries, more attention is paid to their charging capability in short time, therefore it is interesting to investigate how the early-stage non-Markovianity affects the dynamics of a quantum battery.  It would also be benefit for a more accurate assessment of quantum battery performance in the presence of the dissipation.

In this work, we shed light on the impact of the early-stage non-Markovianity on the performance of a charger-mediated quantum battery (CmB) which is modeled by two interacting qubits. One of the qubits acts as the battery and the other acts as the charger which is subjected to a reservoir with the Lorentzian density spectrum.
Through a microscopic derivation we obtain the time-local master equation in Lindblad form that describes the dynamics of the CmB system \cite{lindblad1976,gks1976}. A powerful tool to simulate the dynamics governed by such master equation is the quantum jump method, which considers the stochastic time-evolution of a large number of pure states \cite{dalibard1992prl,molmer1993josab,plenio1998rmp,daley2015advphys}. In the standard quantum jump method, the state of each realization evolves either continuously governed by an effective non-Hermitian Hamiltonian or undergoes normal quantum jumps during the period of positive decay rate.

In the secular region of the considered CmB system, the time-dependent decay rate can be temporarily negative which features the non-Markovianity of the dynamics. In Refs. \cite{piilo2008prl,piilo2009pra}, the negative decay rate is interpreted as the possibility for the system to recover the state before decoherence. The revival of the coherence is realized by the action of the reversed quantum jump that cancels the previous normal quantum jump in the period of positive decay rate. However, when the non-Markovian quantum jump (NMQJ) method is extended to the many-body or driven systems, it is rather challenging due to the dramatically increased trajectory number.

Considerable efforts has been devoted in generalizing the NMQJ method \cite{smirne_PRL2020,luoma_PRL2020,chruscinski_quantum2022,donvil_NC2022,becker_PRL2023,settimo_PRA2024}. The diagrammatic method developed by Chiriac{\'o} {\it et al.} allows to analytically calculate the probability of the trajectories in the NMQJ framework \cite{chiriaco2023prb}. Inspired by this technique, one of the authors has simplified the formalism for the driven-dissipative qubit system \cite{jin2025pra} which is naturally suited for the CmB system presented in Sec. \ref{sec_model}. By combining the Runge-Kutta (RK) method and the NMQJ method, we provide the evidence that the early-stage memory effect may increase the maximal ergotropy at the end of the charging process of the CmB. We also propose a discrete-time scheme for enhancing the performance of the CmB with random local operation in the early-stage evolution.

The paper is organized as follows. In Sec. \ref{sec_model} we demonstrate the model of the CmB system and present the time-local master equation in the Schr{\"o}dinger picture. In Sec. \ref{sec_results}, we investigate the influence of the early-stage memory effects on the performance of the quantum battery. To illustrate the revival of the coherence in the CmB system with negative dephasing rate, we introduce the non-Markovian quantum jump method in Sec. \ref{sec_NMQJ}. A measurement-enhanced quantum battery is proposed in Sec. \ref{sec_MEQB}. We summarize in Sec. \ref{sec_summary}.

\section{The model}
\label{sec_model}
\begin{figure}[htb]
  \includegraphics[width=0.9\linewidth]{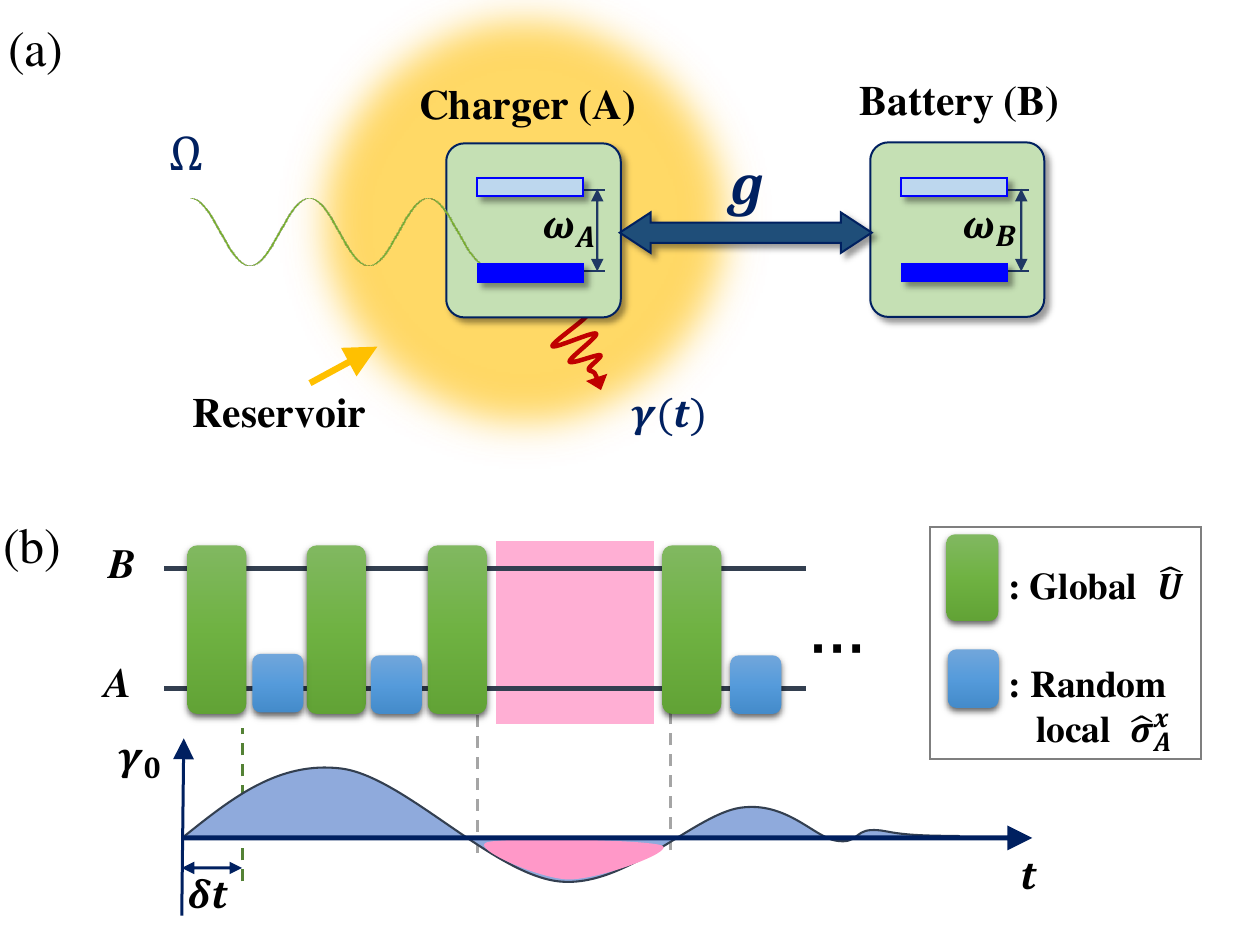}
  \caption{(a) The model. The charger qubit is subjected to the reservoir while the battery qubit is isolated. The spectral densities of the reservoirs in both models are Lorentzian. The charger is driven by an external field as the energy supplier and interacts with the battery. (b) The quantum circuit diagram for the measurement-enhanced quantum battery. The global unitary operations $\hat{U}$ acting on the charger (A) and the battery (B), alternate with the random local operations $\hat{\sigma}^x_A$ acting on the charger (A), in the $\gamma_0>0$ period of the early stage. The local operation is suspended on during the $\gamma_0<0$ period (pink shaded region).}
  \label{fig1_model}
\end{figure}

We consider a CmB system consisting of two qubits with one acting as the charger, denoted by $A$, and the other acting as the quantum battery, denoted by $B$.  The charger is subjected to a thermal bosonic reservoir denoted by $E$, as shown in Fig. \ref{fig1_model}(a). In this scenario, the energy meant to be transferred to the battery is not stored in the charger in advance. Instead it is supplied by the external coherent driving field, or, as will be discussed below, partly by the information backflow from the environment. The charger acts as the mediator between the battery and the energy supplier. This leaves the charger subjected to the environment while the battery isolated from it. Alternatively, the charger can be interpreted as a part of the global environment ($A+E$) of the battery, and it act as the interface with the battery. We would like to notice that the battery itself has the possibility to interact with an additional independent environment, which may induce a detrimental effect by reducing the energy stored in the battery. However, the configuration of the present CmB system allows us to consider this extra dissipation to be negligible.

The free Hamiltonians of the subsystems $A$, $B$ and $E$ are given as follows (set $\hbar=1$ hereinafter),
\begin{equation}
\label{eq_HA}
\hat{H}_A=\frac{\omega_A}{2}\hat{\sigma}_A^z,
\end{equation}
\begin{equation}
\hat{H}_B=\frac{\omega_B}{2}\hat{\sigma}_B^z,
\label{eq_HB}
\end{equation}
and
\begin{equation}
\label{eq_HE}
\hat{H}_E=\sum_{k}{\omega_k\hat{a}_k^\dagger\hat{a}_k},
\end{equation}
where $\hat{\sigma}_X^{\alpha}$ ($\alpha=x,y,z$ and $\text{X}=A,B$) represents the Pauli matrices acting on the subsystem $X$ and $\hat{a}_k$ is the annihilation operator of the $k$-th harmonic oscillator mode of the reservoir. The frequencies $\omega_A$, $\omega_B$ and $\omega_k$ represent the energy separations of the charger, the battery and the energy of the $k$-th bosonic mode, respectively. In addition, the charger is driven by an external field as the energy supply via the following Hamiltonian,
\begin{equation}
\label{eq_Hdr}
\hat{H}_A^{\text{dr}}=\frac{\Omega}{2}(\hat{\sigma}_A^+e^{-i\omega_Lt}+\text{h.c.}),
\end{equation}
where
$\hat{\sigma}^\pm=(\hat{\sigma}^x\pm i\hat{\sigma}^y)/2$ are the corresponding raising and lowering operators. The $\omega_L$ and $\Omega$ is the frequency and the amplitude of the external driving field. Here we denote the detuning of the frequencies between the driving field and the charger as $\Delta =|\omega_A-\omega_L|$.

The Hamiltonians describing the charger-battery and the charger-reservoir interactions are given as follows,
\begin{equation}
\label{eq_HAB}
\hat{H}_{AB}=g(\hat{\sigma}_A^+\hat{\sigma}_B^-+\text{h.c.}),
\end{equation}
\begin{equation}
\label{eq_HAE}
\hat{H}_{AE}=\sum_{k}{\left(g_k\hat{\sigma}_A^+\hat{a}_k+\text{h.c.}\right)},
\end{equation}
where $g$ and $g_k$ are the coupling strength of charger with the battery, and with the $k$-th bosonic mode, respectively. The operators are given in the computational basis, i.e. an orthogonal basis $\{\ket{\uparrow},\ket{\downarrow}\}$, determined by the upper and lower levels of the qubit. Making use of the spectral density defined by $J(\omega)=\sum_{k}{|g_k|^2\delta(\omega-\omega_k)}$, one has $\sum_{k}{|g_k|^2}\rightarrow\int \text{d}\omega J(\omega)$ in the limit of a continuum of the reservoir modes. In this work we focus on the reservoir with a Lorentzian spectral density,
\begin{equation}
J_{\text{Lor}}(\omega) = \frac{\eta^2}{2\pi}\frac{\lambda^2}{(\omega-\omega_0)^2+\lambda^2},
\label{eq_Lorentzian}
\end{equation}
where $\omega_0$ is the central frequency, $\lambda$ is the width of the spectrum and $\eta^2$ denotes the effective coupling constant.

In the limit of weak coupling, one can obtain the time-local master equation describing the time-evolution of the joint state of the charger and the battery as follows,
\begin{equation}
\dot{\rho}_{AB}(t)= -i[\hat{H}_0+\hat{H}_1,\rho_{AB}(t)] + {\cal D}[\rho_{AB}(t)],
\label{eq_loc_time_ME}
\end{equation}
where $\hat{H}_0=\frac{\Omega}{2}\hat{\sigma}_A^x+\frac{\omega_B}{2}\hat{\sigma}^z_B$ and $\hat{H}_1=g(e^{i\omega_Lt}\hat{\sigma}_A^+\hat{\sigma}_B^-+\text{h.c.})$. As detailed in the Appendix \ref{sec_app_A}, Eq. (\ref{eq_loc_time_ME}) is derived in the rotating frame with $e^{-i\omega_L\hat{\sigma}_A^z t/2}$ and under the resonant condition, i.e. $\omega_A=\omega_B=\omega_L$. The first term of the r.h.s. of Eq. (\ref{eq_loc_time_ME}) describes the unitary dynamics governed by $\hat{H}_0+\hat{H}_1$. While the second term is the dissipator describing the dissipation induced by the reservoir and, in general, containing the secular and nonsecular terms \cite{haikka2010physcri}. Because the quantum battery is isolated from the reservoir, the dissipator acts locally on the charger.

In order to specify the dissipator ${\cal D}$ in Eq. (\ref{eq_loc_time_ME}), two parameters are introduced as follows,
\begin{equation}
p=\frac{\tau_C}{\tau_A}=\frac{\sqrt{\Delta^2+\Omega^2}}{\lambda},
\label{eq_parameter_p}
\end{equation}
\begin{equation}
s=\frac{\omega_0-\omega_L}{\lambda},
\end{equation}
where  $\tau_C=\lambda^{-1}$ characterizes the time scale of the reservoir correlation and $\tau_A=(\Delta^2+\Omega^2)^{-1/2}$ characterizes the typical time of the charger over which its state varies appreciably. The former parameter $p$ identifies the region in which the secular approximation is valid, while the latter parameter $s$ characterizes the detuning of the central frequency of the Lorentzian spectrum from the frequency of the driving field in units of $\lambda$.

\begin{figure}[hbt]
  \includegraphics[width=1\linewidth]{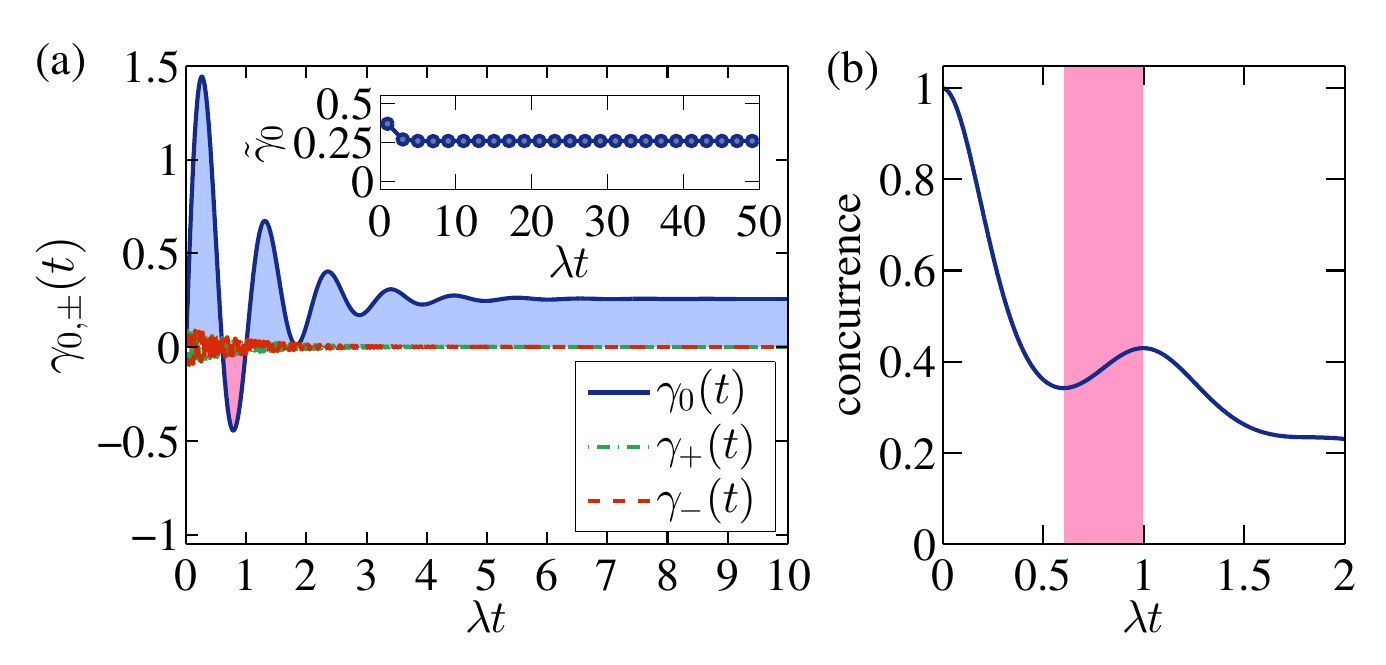}
  \caption{(a) The time-dependence of the decay rates of the charger. The solid line denotes the behavior of $\gamma_0$. In the early-stage $\lambda t\lesssim 4$, the positive $\gamma_0$ is punctuated by a segment with negative $\gamma_0$. For $\lambda t\gtrsim 4$, the $\gamma_0$ remains positive and reach to the asymptotic value at large $\lambda t$. The decay rates $\gamma_+(t)$ and $\gamma_-(t)$ oscillates between the positive and negative values. The amplitude is so small ($<10^{-2}$) compared with $\gamma_0$ that can be neglected. The inset shows the coarse-grained value of $\gamma_0$. The parameters are chosen as $p=100$ and $s=6$. (b) The concurrence of the CmB system with $g=0$. The initial state is the maximally entangled state. A revival of entanglement can be observed during the period of negative $\gamma_0$ indicating the back flow of information. The shaded regions in red indicate the period of negative $\gamma_0$.}
  \label{fig2_decayrate}
\end{figure}

Since the goal of this work is to highlight the non-Markovian effects on the performance of the CmB, we will focus on the case of
$p\gg1$ for which $\tau_A$ is much shorter than $\tau_C$. In such a situation, the nonsecular parts in ${\cal D}$, which is shown in Eq. (\ref{eq_A_nonsecular}), oscillate rapidly with respect to the eigen frequency $\omega=\sqrt{\Delta^2+\Omega^2}$ of the charger, and thus can be neglected  \cite{piilo2009pra,haikka2010physcri}. Moreover, if the typical frequency of the charger $\omega_A$ is far detuned from the cental frequency of the reservoir $\omega_0$, i.e. $s\gg1$, the decay rates for the incoherent loss and gain (denoted by $\gamma_{\pm}$) in the dissipator are suppressed. Therefore the dissipator ${\cal D}$ is reduced to the following
\begin{equation}
{\cal D}[\rho_{AB}]=\gamma_0(t)(\hat{\sigma}_A^x\rho_{AB}\hat{\sigma}_{A}^x - \rho_{AB}),
\label{eq_dissipator}
\end{equation}
where the time-dependent decay rate is determined by the real part of the reservoir correlation function as described in Appendix A, and is given by
\begin{equation}
\gamma_0(t)=\frac{\eta^2\{1-e^{-\lambda t}[\cos{(s\lambda t)}-s\sin{(s\lambda t)}]\}}{4(1+s^2)}.
\label{eq_gamma_t}
\end{equation}

In Fig. 2(a) it is shown the decay rates in the secular region with $p=100$ and $s=6$. One can see that (i) the amplitudes of the $\gamma_{\pm}$ is much smaller than $\gamma_0$ indicating that the loss and gain processes are strongly suppressed; (ii) the $\gamma_0$ oscillates at the early stage and then asymptotically approaches to a positive constant $\gamma_0(\infty)$ ; and (iii) the $\gamma_0$ can be temporally negative during an intermediate period $0.6\lesssim \lambda t\lesssim 1$.
The positive decay rate indicates that the time-evolution of the composite system of the charger and the quantum battery is Markovian. On the contrary, the negative decay rate violates the complete positive and trace-preserving of the quantum master and may give rise to the possibility for the system to recover the state before decoherence.

When the scale of the time resolution is large, the decay rate can be coarse-grained as $\tilde{\gamma}_0(T)=\frac{1}{T}\int_{t-T/2}^{t+T/2}{\gamma_0(t)dt}$ where $T$ is the coarsening size. As shown in the inset of Fig. \ref{fig2_decayrate}, the coarse-grained decay rate $\tilde{\gamma}_0$ with $T = \lambda t$ smooths out the non-Markovianity in the early-stage and the dynamics can be regarded as a Markovian process. Indeed the coarse-graining treatment is a good approximation for the case that the typical time scale of the system of interest is larger than $T$, or for investigating the steady-state property.
However, in the present scenario we assume that for $t<0$, $A$, $B$, and $E$ do not interact and are isolated from each other.  At time $t=0$, the interactions between $A-B$ and $A-E$ are switched on and the energy is transferred to the battery through the charger, which is the so-called charging process. We define the charging time $t_{\text{ch}}$ as the time at which the ergotropy of the battery reaches the maximum. From the perspective of a practical application scenario, the user may expect the charger time to be as short as possible. In this case, one focuses on the temporal behavior of the CmB system in the early stage and the details of $\gamma_0(t)$ should be taken into account. We will discuss the impact of the early-stage behavior of $\gamma_0$ in the next section.

\section{Performance of the quantum battery}
\label{sec_results}

Let us start by showing the memory effect during the period of negative decay rate by simulating the non-Markovian dynamics described by Eq. (\ref{eq_loc_time_ME}). We perform a direct integration of the master equation by using the standard fourth-order RK method with the CmB system initialized in the maximally entangled state $|\psi\rangle_{AB}=\frac{1}{\sqrt{2}}(|\uparrow^z_A\downarrow^z_B\rangle+|\downarrow^z_A\uparrow^z_B\rangle)$, where $|\uparrow^\alpha_X\rangle$ and $|\downarrow^\alpha_X\rangle$ are the eigenvectors of $\hat{\sigma}_X^\alpha$ with the eigenvalues $+1$ and $-1$, respectively. The entanglement is measured by the concurrence \cite{wootters1998prl}. For a two-qubit system in state $\rho$, the concurrence is given by $C(\rho)=\max{\{0,\sqrt{\lambda_1}-\sqrt{\lambda_2}-\sqrt{\lambda_3}-\sqrt{\lambda_4}\}}$ with $\lambda_1\ge\lambda_2\ge\lambda_3\ge\lambda_4$ being the eigenvalues of the matrix $\rho(\hat{\sigma}^y\otimes\hat{\sigma}^y)\rho^*(\hat{\sigma}^y\otimes\hat{\sigma}^y)$.

Since the dissipator acts locally on the charger, the entanglement is expected to decrease monotonically in the Markovian dynamics. The time-evolution of the entanglement is shown in Fig. \ref{fig2_decayrate}(b). A revival of entanglement can be observed in the period of $\gamma_0<0$ signaling the backflow of information from the reservoir the system which is a feature of the non-Markovianity.

Now we investigate the performance of the CmB system with dephasing. We will focus on the charging process for an empty battery in the empty state $\ket{\downarrow^z_B}$. At the end of the charging process, the energy contained in the battery, by convention, measured by the expectation value of the population of the upper level is given as follows,
\begin{equation}
E_B(t)=\frac{\omega_B}{2}\text{tr}[(\hat{\sigma}^z_B+\hat{\mathbb{I}}_B)\rho_{AB}(t)],
\end{equation}
where $\hat{\mathbb{I}}_B$ is the identity matrix. Another figure of merit of the quantum battery is the ergotropy which quantifies the maximal extractable work under a cyclic unitary transformation $U$,
\begin{equation}
{\cal E}_B(t)=E_B(t)-\min_{U\in {\cal U}_c}{\{\text{tr}[\hat{H}_B\hat{U}_B\rho_B(t)\hat{U}_B^\dagger]\}},
\end{equation}
where $\rho_B(t)$ is the reduced density matrix of the battery and ${\cal U}_c$ is the set of cyclic unitary operations \cite{allahverdyan2004epl,shi2022prl}. For a qubit battery with the free Hamiltonian (\ref{eq_HB}), the mean-energy and the ergotropy can be obtained explicitly as follows,
\begin{equation}
E_B(\tau)=\frac{\omega_B}{2}[1+r_z(\tau)],
\end{equation}
\begin{equation}
{\cal E}_B(\tau)=\frac{\omega_B}{2}[r(\tau)+r_z(\tau)],
\end{equation}
where $r(\tau)$ and $r_z(\tau)$ are the modulus and the $z$ component of the Bloch vector associated with the $\rho_B(\tau)$ \cite{farina2019prb}.

\begin{figure}[hbt]
  \includegraphics[width=1\linewidth]{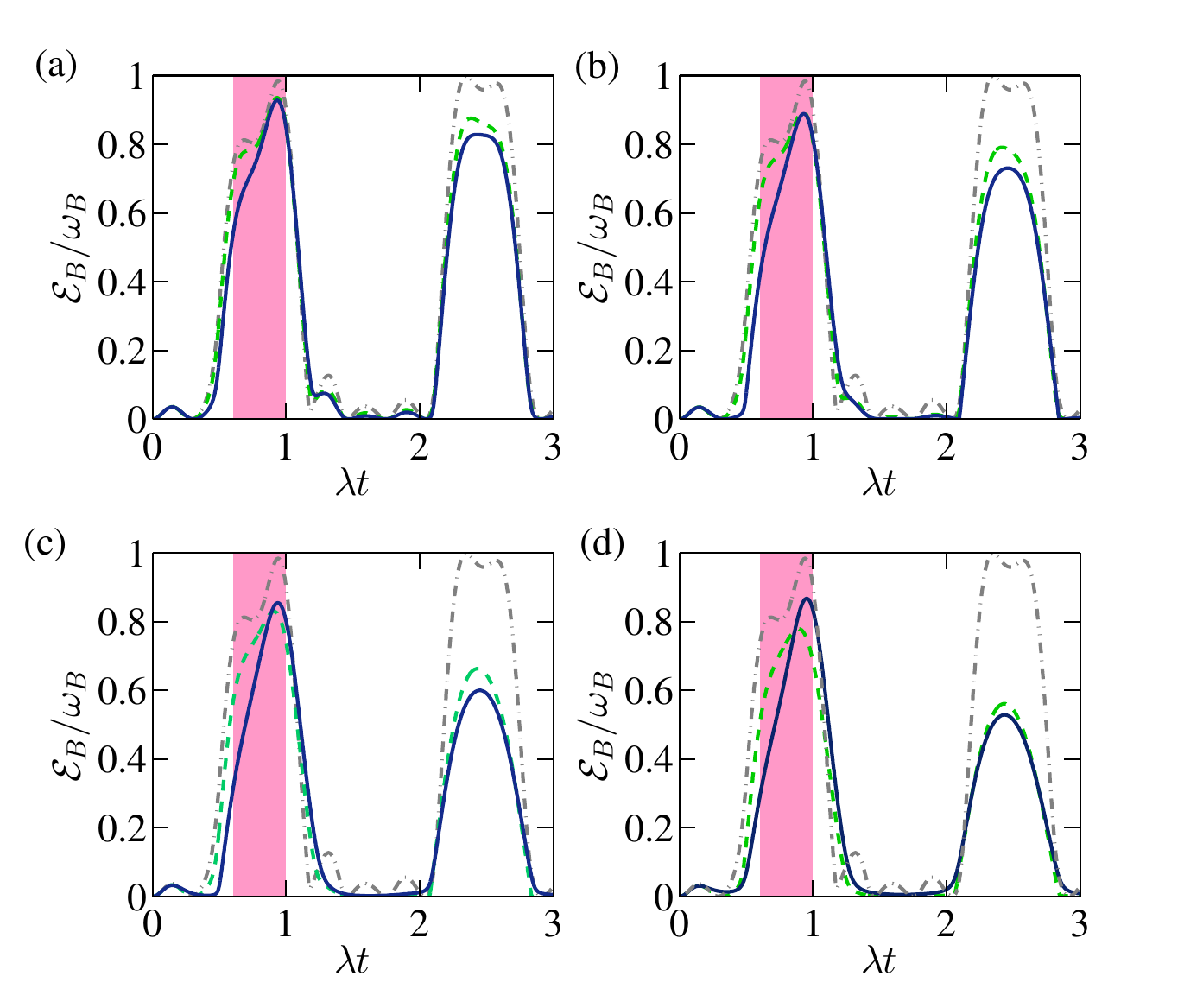}
  \caption{The time-evolution of the ergotropy of the charger-mediated battery with $\omega_A=\omega_B=\omega_L=100\lambda$, $\Omega=-0.1\omega_A$ and $g=4$. The coupling strength between the charger and the reservoir in Eq. (\ref{eq_gamma_t}) are $\eta^2=0.5$, $1$, $2$ and $3$ for panels (a)-(d). The initial state is $\ket{\psi_{AB}(0)}=\ket{\uparrow^y_A\downarrow^z_B}$. The solid, dashed, and dash-dotted lines represents the cases (i), (ii), and (iii), respectively. The shaded region in red indicates the period of $\gamma_0<0$. }
  \label{fig3_ergotropy}
\end{figure}

For a concrete example, we specialize the parameters in Eqs. (\ref{eq_HA})-(\ref{eq_HAB}) as $\omega_A=\omega_B=\omega_L=100\lambda$ and $\Omega = -0.1\omega_A$. The chosen parameters satisfy both the resonant condition and the requirement of secular approximation, and thus produce the decay rate $\gamma_0(t)$ in Fig. \ref{fig2_decayrate}(a). In order to uncover the early-stage memory effect on the performance of the quantum battery, we investigate dynamics for three cases: (i) the non-Markovian dephasing with time-dependent decay in Eq. (\ref{eq_gamma_t}), (ii) the Markovian dephasing with a constant decay rate which is the asymptotic value of Eq. (\ref{eq_gamma_t}), i.e. $\gamma_0(\infty)$, and (iii) the unitary evolution without dephasing.

The dynamics of the ergotropy for various coupling strength between the charger and the reservoir are shown in Fig. \ref{fig3_ergotropy}. The behavior of the ergotropy under unitary time evolution shows the battery is maximally charged for the first time at $\lambda t\approx0.95$, referred to as the charging time. The charging time in case (iii) is determined by the coupling strength $g$ in Eq. (\ref{eq_HAB}) and could serve as a benchmarker for cases (i) and (ii).

From Figs. \ref{fig3_ergotropy}(a) to \ref{fig3_ergotropy}(d), one can see that the peak of the ergotropy in cases (i) and (ii) is lowered due to the dephasing, and the height of the peak decreases as $\eta$ increases. To be specific, the charging time in case (ii) is shorten as the $\eta$ increases but at a cost of the decreasing of the maximum of the ergotropy. Comparing the case (i) and (ii), the charging time in case (ii) is slightly shorter than that in case (i). More interestingly, the maximal ergotropy in case (i) becomes larger than that of case (ii) as the $\eta$ increases. A more detailed explanation can be found in Appendix. \ref{sec_app_B}. This can be understood as the following. In case (i), before the period of $\gamma_0<0$, the temporal magnitude $|\gamma_0(t)|$ oscillates around $\gamma_0(\infty)$ resulting in a stronger decoherence and thus a lower ergotropy compared to case (ii). While during the period of $\gamma_0(t)<0$ (the shaded region in the panels), the information, previously lost into the reservoir, flows back to the CmB system to recover the behavior of case (iii) manifested by a rapid growth in (the steep slope of) the ergotropy. On the other hand, the large magnitude of the negative $\gamma_0$ indicates a strong memory effect, by means of more likely to implement the reversed quantum jump as will be seen soon, therefore the maximal ergotropy is close to that of case (iii) and higher than that of case (ii).

\section{Non-Markovian quantum jump}
\label{sec_NMQJ}
The memory effect during the period of negative $\gamma_0(t)$ can be understood in the unraveling of the time-local master equation (\ref{eq_loc_time_ME}) with the NMQJ method. In the quantum jump method, Eq. (\ref{eq_loc_time_ME}) can be recast as follows,
\begin{equation}
\dot{\rho}_{AB}(t)=-i[\hat{H}_{\text{eff}}\rho_{AB}(t)-\rho_{AB}(t)\hat{H}_{\text{eff}}^\dagger]+\gamma_0(t)\hat{\sigma}_A^x\rho_{AB}(t)\hat{\sigma^x_A},
\label{eq_ME_eff}
\end{equation}
where the non-Hermitian effective Hamiltonian is $\hat{H}_{\text{eff}}=\hat{H}_0+\hat{H}_1-i\gamma_0(t)\hat{\mathbb{I}}/2$. An initial pure state evolves either continuously under the non-unitary operator $\exp{(-i\hat{H}_{\text{eff}}t)}$ or undergoes a discontinuous operation via the action of $\hat{\sigma}_A^x$.
The density matrix in Eq. (\ref{eq_loc_time_ME}) can thus be represented as the following incoherent mixture of all the possible states of the members in the resulting ensemble,
\begin{equation}
\rho(t) = \sum_n{K_n^\alpha(t)|\psi_n^\alpha(t)\rangle\langle\psi_n^\alpha(t)|},
\label{eq_rhot_ensemble}
\end{equation}
where $\ket{\psi_n^\alpha(t)}$ denotes one of the possible states at time $t$ and $K_n^\alpha(t)$ is the corresponding weight.

In the NMQJ method, the time axis is discretized by the time interval $\delta t$. In Ref. \cite{jin2025pra}, different states $\ket{\psi_n^\alpha}$ are considered to originate from different quantum trajectories which are labeled by $H_n^\alpha$ with $\alpha$, representing the time sequence, is an array of time that records the time at which the system undergoes a quantum jump. The number of elements in the array $\alpha$ is denoted by $n=\#\alpha$, representing the number of quantum jumps that the system has undergone.
Here the quantum state refers to the CmB system and the subscript `AB' has been omitted. Such {\it trajectory} describes the continuous time evolution of the system's state governed by the effective non-Hermitian Hamiltonian starting from the time point $t=\alpha(n)$. Specially, the quantum state associated to the no-jump trajectory $H_0^{\o}$ at time $t_s$ is given by
\begin{equation}
|\psi_0^{\o}(t_s)\rangle=\frac{\exp{(-i\hat{H}_{\text{eff}}t_s)}|\psi(t_0)\rangle}{||\exp{(-i\hat{H}_{\text{eff}}t_s)}|\psi(t_0)\rangle||},
\end{equation}
and the quantum state associated to a generic quantum trajectory $H_n^\alpha$ at $t_s>t_w$ is given by the following recurrence relation
\begin{equation}
|\psi_n^{\alpha}(t_s)\rangle=\frac{\exp{[-i\hat{H}_{\text{eff}}(t_s-t_w)]}|\psi_n^\alpha(t_w)\rangle}{||\exp{[-i\hat{H}_{\text{eff}}(t_s-t_w)]}|\psi_n^\alpha(t_w)\rangle||},
\end{equation}
where $|\psi_n^\alpha(t_w)\rangle=\hat{\sigma}_A^x|\psi_{n-1}^{\alpha'}(t_w)\rangle/||\hat{\sigma}_A^x|\psi_{n-1}^{\alpha'}(t_w)\rangle||$, and $\alpha=[\alpha',t_w]$ with $\alpha'$ being the time sequence of the trajectory $H_{n-1}^{\alpha'}$ which is the mother trajectory of $H_n^\alpha$. Namely, the trajectory $H_{n-1}^{\alpha'}$ undergoes a quantum jump at $t_w$ and consequently generates the trajectory $H_n^\alpha$.

To incorporate the memory effect during the period of $\gamma_0<0$, the set of trajectories $H_n^\alpha$ characterized by the time sequences satisfying $\alpha=[\alpha',t_w]$ are classified into the {\it trajectory class} represented by $\{H_n^\alpha\}$. The $H_n^\alpha$s sharing the same $\alpha'$ in $\alpha$ indicates that these trajectories originate from the same mother trajectory but under the action of a quantum jump at different time $t_w$. Assuming that the memory time is infinitely long, the trajectories belong to the $\{H_n^\alpha\}$ will be brought back to the same quantum state by the reversed quantum jump.

The $K_n^\alpha$, referred to as the {\it existence probability}, represents the weight of the realizations staying in the state $|\psi_n^\alpha(t_s)\rangle$ in the limit of infinite number of realizations. Apparently, the sum of all the existence probability should be one by definition. The details in deriving the explicit expression of
$K_n^\alpha$ can be found in Ref. \cite{jin2025pra} and we apply the result to the present model.

Since the jump operator in Eq. (\ref{eq_loc_time_ME}) is unitary, i.e. $(\hat{\sigma}^x_A)^2=\mathbb{I}$, the probability of a normal quantum jump in the regime of positive $\gamma_0(t)$ is given by $p_n^\alpha(t) = \gamma_0(t)\delta t$ which is state-independent. Therefore the existence probability of the no-jump trajectory yields
\begin{equation}
K_0^{\o}(t_w)=\prod_{s=1}^{w-1}\left[1-\gamma_0(t_s)\delta t\right].
\label{eq_expr_K0}
\end{equation}

By virtue of Eq. (\ref{eq_expr_K0}), the existence probability $K_n^{\alpha}$ of a generic trajectory with $n\ge1$ at the end of the $\gamma_0>0$ period, i.e. the moment $t =t_P$, can be obtained iteratively as follows,
\begin{equation}
K_n^{\alpha}(t_P)=K_{n-1}^{\alpha'}(t_{w-1})\gamma_0(t_w)\delta t\prod_{s=w+1}^{P}\left[1-\gamma_0(t_s)\delta t\right].
\end{equation}
The survival of $H_n^\alpha$ at $t_P$ is a consequence of quantum trajectory $H_{n-1}^{\alpha'}$ undergoes a quantum jump at time $t_w$ followed by no more jumps until $t_P$.

In the period of $\gamma_0<0$, the normal quantum jump is suspended while the {\it reversed} quantum jump is switched on. The former prohibits the creation of new quantum trajectory and the latter allows the existing quantum trajectory to jump back to its mother trajectory. The existence probability of $K_0^{\o}$ shares the same form as Eq. (\ref{eq_expr_K0}) for $\gamma_0<0$, therefore the coherence is restored as $K_0^{\o}$ increases. For a generic trajectory, after the first time interval of the $\gamma(t)<0$ period, the existence probability of $H_n^{\alpha}$ is given as follows,
\begin{eqnarray}
K_n^{\alpha}(t_P+\delta t)&=&K_n^\alpha(t_P)-K_n^\alpha(t_P)\gamma_0(t_P)\delta t\cr\cr
&&+\frac{K_{n-1}^{\alpha'}(t_P)K_n^\alpha(t_P)}{\sum_{\alpha}{K_n^{\alpha}(t_P)}}\gamma_0(t_P)\delta t.
\label{eq_Kn_alpha}
\end{eqnarray}
The second term, $-K_n^\alpha(t_P)\gamma_0(t_P)\delta t$, on the right-hand side of Eq. (\ref{eq_Kn_alpha}) represents the increasing of $K_n^\alpha$ due to the reversed jump from the subtrajectories of $H_n^\alpha$. The third term, $\frac{K_{n-1}^{\alpha'}(t_P)K_n^\alpha(t_P)}{\sum_{\alpha}{K_n^{\alpha}(t_P)}}\gamma_0(t_P)\delta t$, on the right-hand side of Eq. (\ref{eq_Kn_alpha}) represents the decreasing of $K_n^\alpha$ due to the reversed jump from $K_n^\alpha$ to its mother trajectory $H_{n-1}^{\alpha'}$, with $\alpha=[\alpha', t_w]$ \cite{cf_jin2025pra}.

\begin{figure}[thb]
 \includegraphics[width=1\linewidth]{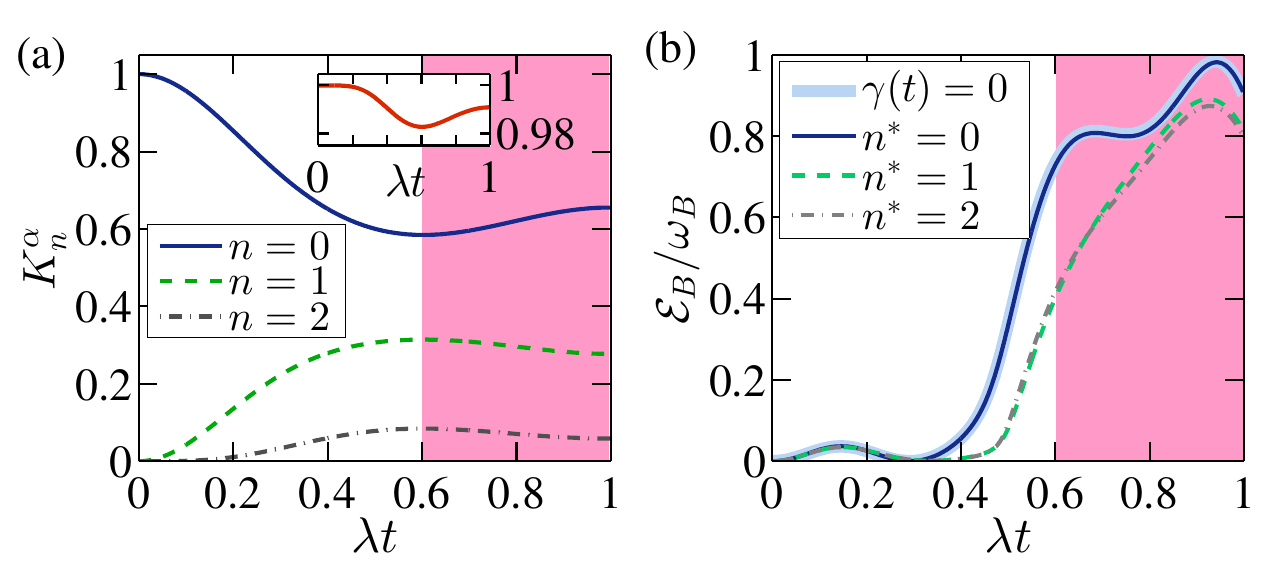}
 \caption{(a) The existence probabilities of the no-jump, one-jump and two-jump trajectory classes. During the period of positive decay rate the $K_o^{\o}$ decreases due the actions of the normal jumps while the $H_1^\alpha$ and $H_2^\alpha$ trajectories are created. The inset shows the sum of the existence probabilities up to $n=2$ which is close to one throughout the time window. (b) The thick line represents the ergotropy under unitary time evolution. The solid, dashed and dash-dotted (thin) lines represent the ergotropy with respect to the density matrix computed by the NMQJ method with  the summation in Eq. (\ref{eq_rhot_ensemble}) up to $n^*=0$, $1$, and $2$, respectively. The initial state is $\ket{\uparrow_A^y\downarrow_B^z}$. The coupling between the charger and the battery is $g=4$ and the time interval is $\delta t=5\times10^{-4}\lambda^{-1}$. Other parameters are chosen the same as in Fig. \ref{fig3_ergotropy}. }
\label{fig4}
\end{figure}

In Fig. \ref{fig4}(a) it is shown the time-evolution of the existence probabilities with various $n$. In the $\gamma_0>0$ period, the $K_0^{\o}$ decreases while $K_1^\alpha$ and $K_2^\alpha$ increase as the normal jumps take place. In contrast, in the $\gamma_0<0$ period, a revival of $K_0^{\o}$ can be observed due to the contribution from the $K_1^\alpha$ and $K_2^\alpha$. Moreover, as shown in the inset of Fig. \ref{fig4}(a), the sum $\sum_{n=0}^2{K_n^\alpha}$ also exhibits a revival in the $\gamma_0<0$ period and the magnitude is larger than $0.98$ throughout the time window. Therefore, for the decay rate presented in Fig. \ref{fig2_decayrate}(a), the sum in Eq. (\ref{eq_rhot_ensemble}) can be truncated at $n=2$ to construct the density matrix.

In order to reveal the contribution of each trajectory class in constructing the density matrix, we compare the ergotropies of the quantum battery with respect to the pure state under unitary time evolution [case(iii), $\gamma_0=0$] and to the state of the no-jump trajectory [case(i) with the sum in Eq. (\ref{eq_rhot_ensemble}) truncated at $n^*=0$]. As shown by the thick and thin solid lines in Fig. \ref{fig4}(b), these two results coincide. This is because the identity matrix in the imaginary part of $\hat{H}_{\text{eff}}$ only introduces a global coefficient to $\ket{\psi_0^{\o}}$ which is proportional to $\gamma_0$. The impact of the dephasing manifests in trajectories that have undergone at least one quantum jump. The ergotropy computed by the density matrix in Eq. (\ref{eq_rhot_ensemble}) with the summation up to $n^*=1$ and $2$ are shown by the dashed and dash-dotted lines in Fig. \ref{fig4}(b). One can see that the ergotropy converges at $n^*=2$ indicating the contributions from the quantum trajectories that undergo more than two quantum jump can be neglected. Moreover, for $0.4\lesssim\lambda t\lesssim0.6$ the ergotropy in the dephased time evolution deviates from that in the unitary evolution due to the creation of the trajectories belong to ${H_n^\alpha}$ and ${H_2^\alpha}$. When entering into the region of $\gamma_0<0$, the reversed quantum jump brings the $H_n^\alpha$ and $H_2^\alpha$ trajectories back to the no-jump trajectory. Therefore the ergotropy under dephasing evolution becomes closer to that under unitary evolution because the history of dephasing being (partly) cancelled. At this sense, the negative $\gamma_0$ can be considered as a signal of the memory effect.

\begin{figure}[tbh]
 \includegraphics[width=1\linewidth]{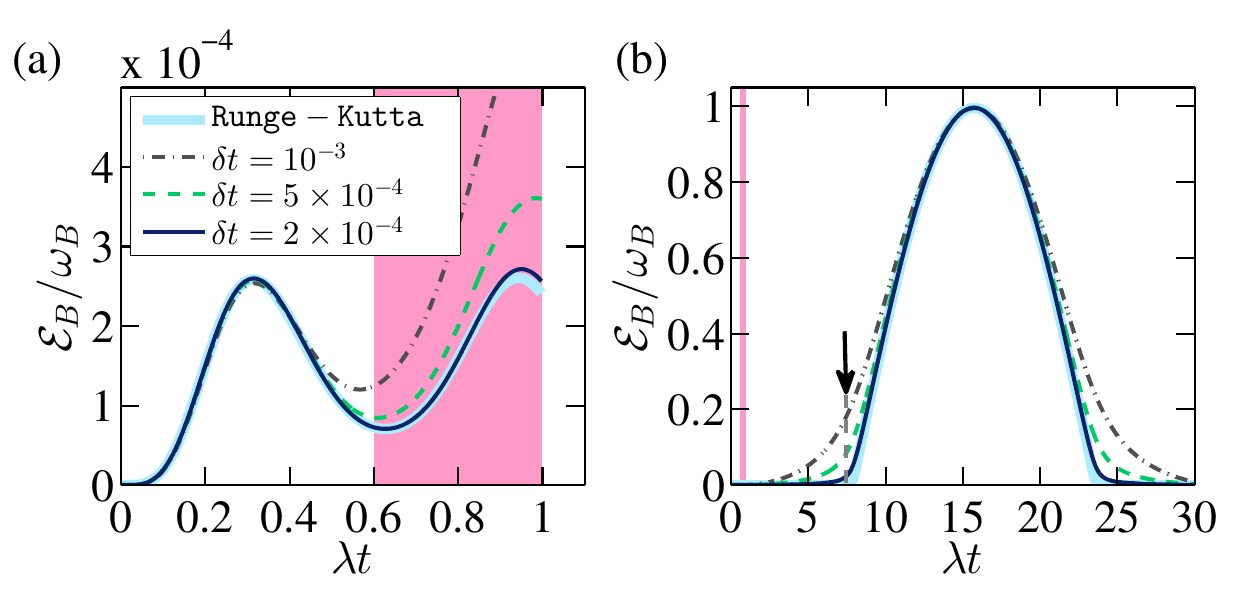}
 \caption{(a) The early-stage time evolution of the ergotropy. The thick solid line represents the results computed by the fourth-order Runge-Kutta method, while the thin solid, dashed, and dot-dashed line represent the NMQJ results with time interval $\delta t=2\times10^{-4}\lambda^{-1}$, $5\times10^{-4}\lambda^{-1}$ and $10^{-3}\lambda^{-1}$, respectively. (b) The long-time evolution of the ergotropy. The initial state is $|\psi_{AB}(0)\rangle=|\downarrow_A^z\downarrow_B^z\rangle$. The parameters are chosen as $g=0.2$, $\eta^2=1$ and the others are the same as in Fig. \ref{fig3_ergotropy}.}
\label{fig5}
\end{figure}

\section{Measurement-enhanced quantum battery}
\label{sec_MEQB}
The implementation of the NMQJ method assumes that time interval $\delta t$ is so short that only one quantum jump takes place during $\delta t$. To verify, the comparison of the ergotropies computed by the fourth-order RK method and by the NMQJ method with various $\delta t$ is shown in Fig. \ref{fig5}(a). To highlight the effects of the non-Markovianity, the case of weak coupling between the charger and the battery is considered in which the early-stage behavior is mainly governed by the dephasing.

One can see that the NMQJ result agrees with the RK result when the time interval is chosen as $\delta t=2\times10^{-4}\lambda^{-1}$. A large $\delta t$ causes the state of the quantum battery to deviate from the true one at the end of the $\gamma<0$ region. Although the results obtained with larger $\delta t$ are incorrect concerning the solution to Eq. (\ref{eq_ME_eff}), it is still worthy to investigate the subsequent dynamics and, as be illustrated below, may motivate a scenario to enhance the performance of quantum battery.

Recall that in the NMQJ implementation, the physical meaning of $\delta t$ is the time window over which the system either evolves continuously or undergoes an action of local coherent flip on the charger. The probability is essentially determined by the time-dependent $\gamma_0$.
In the limit of $\delta t\rightarrow0$, the NMQJ method precisely simulate the dynamics described by the master equation (\ref{eq_loc_time_ME}). This motivates us to design a quantum circuit acting the CmB system as shown in Fig. \ref{fig1_model}(b). It is composed of the global unitary operation $\hat{U}_{AB}$ and the probabilistic local coherent spin flip operation $\hat{\sigma}^x_A$. The global operation is generated by the Hamiltonian $\hat{H}_0+\hat{H}_1$ over an evolution time $\delta t$, i.e. $\hat{U}=\exp{[-i(\hat{H}_0+\hat{H}_1)\delta t]}$. From Fig. \ref{fig1_model}(b), one can see that the global unitary evolution is punctuated by the local operations which play the role of the quantum jumps. For a certain period of $\gamma_0>0$, increase the time interval $\delta t$ thus means a reduction in the number of local operations (quantum jumps). The probability whether to implement the local $\hat{\sigma}^x_A$ or not is determined by $\gamma_0(t)\delta t$. During the $\gamma_0<0$ period, the $\gamma_0(t)\delta t$ is negative and is not a physical probability. It can be interpreted as that in the $\gamma_0<0$ period the normal quantum jump is suspended, or namely, the local operation should be absent in the circuit.

We take the temporal states at the end of the $\gamma_0<0$ region for each $\delta t$ to continue the time evolution in the subsequent $\gamma_0>0$ region. The long-time behaviors are shown in Fig. \ref{fig5}(b). One can see that the time evolution of the ergotropy with a history of larger $\delta t$ exhibits an advantage over the others. On the one hand, as indicated by the vertical arrow, the quantum battery, which allows only one quantum jump within $\delta t=10^{-3}\lambda^{-1}$ in the early stage, has been charged to twenty percent ergotropy at $\lambda t\approx7.5$, shown by the dash-dot line. However, at $\lambda t\approx7.5$ the ergotropies for other choices of $\delta t$ are much less. This means that a larger $\delta t$ facilitates a rapid increase of the ergotropy in the early stage of the charging process. On the other hand, the ergotropy curve of $\delta t=10^{-3}\lambda^{-1}$ remains above the curves for small $\delta t$, including the case of unitary evolution, until it reaches the maximum at $t_{\text{ch}}\approx15.5\lambda^{-1}$. Therefore, the local operations, i.e. local quantum jump on the charger, in the early stage are beneficial for improving the charging characteristic curve of the ergotropy. In this sense, we may refer the circuit diagram in Fig.  \ref{fig1_model}(b) as a scenario of measurement-enhanced quantum battery.

\section{summary}
\label{sec_summary}
In summary, we have demonstrated the early-stage memory effect on the performance of the dephasing CmB system consisted of two qubits. From a microscopic derivation, we obtained the time-local master equation in Lindblad form with local dephasing on the charger qubit. The time-dependent dephasing rate $\gamma_0(t)$ exhibit an oscillation at the early time and then approaches to the asymptotic value $\gamma_0(\infty)$ in the late time. The temporal negative $\gamma_0$ violates the complete positivity of the master equation and leading to a non-Markovianity of the dynamics of the CmB system. We found that the strong non-Markovianity is beneficial to the performance of the quantum battery. In comparison with the Markovian CmB system with the constant dephasing rate $\gamma_0(\infty)$, the non-Markovian one increases the maximal ergotropy at the end of the charging process. This is because the memory effect during the $\gamma_0<0$ region tends to recover the performance of the battery under unitary evolution.

In order to highlight the memory effect, we adopted the NMQJ method to unravel the Lindblad master equation. We found that the time evolution of the ergotropy with respect to the no-jump state exhibits the same behavior as that in the unitary evolution. The deviation of the ergotropy in the dephased scenario is due to the contributions from the trajectory classes with at least one quantum jump. Actually, we show that a truncation at $n=2$ is sufficient for the sum to construct the density matrix. The reversed jump in the period of $\gamma_0<0$ period transforms the $H_1^\alpha$ and $H_2^\alpha$ trajectories to the no-jump trajectory level by level. Therefore the information lost in earlier time is recovered and the maximal ergotropy could exceed the one in the process under Markovian approximation.

The time interval is essential in the simulation by the NMQJ method. We show that a larger time interval may lead the numerical result to deviate from the true value. Because the number of quantum jumps are restricted so that the continuous time evolution cannot be effectively simulated. Moreover the deviations remain in the subsequent time-evolution and is responsible for an enhancement of the battery performance. This motivates us to propose a discrete time measurement-enhanced scheme at the early-stage of the charging process. A quantum circuit version of such a scheme is proposed. With a relatively larger time interval, the state at the end of the non-Markovian region may lead to an improvement of the charging speed.

Finally, we note that our discussion is concentrated on the bosonic reservoir with Lorentzian spectral density. The spectral density plays significant role in determining the time-dependent decay rates in the time-local master equation. For example, for the Ohmic reservoir, in the secular regime the decay rates $\gamma_\xi(t)$ are of similar order of magnitude, indicating that the quantum jumps and dephasing rate contributes to the dynamics of system \cite{haikka2010}. Therefore investigating the combined effects of the multiple non-Markovian decay channels on the performance of the CmB system would be interesting for the future work. Additionally, although studies on the performance of the quantum battery with the fermionic reservoir are relatively few, it is an intriguing direction worth exploring \cite{chang1985prb}. Along this line, the fermionic quantum state diffusion equation provides a suitable approach \cite{jing2010prl,zhao2012pra}. For investigating the non-Markovian effects on the quantum battery in many-body settings, although the NMQJ method may face certain limitations, the {\it collisional model} can offer a promising approach, especially for the setups with boundary dissipation \cite{morrone2023qst,YL2020pra,YL2023pra}.

\section*{ACKNOWLEDGMENTS}
J.J. thanks Prof. H. Z. Shen and Prof. X. Y. Zhao for valuable discussion. This work is supported by Natural Science Foundation of Liaoning Province of China No. 2025-MS-009.

\begin{appendix}
\section{Microscopic derivation of the master equation}
\label{sec_app_A}
In the Scr{\"o}dinger picture, the Hamiltonian of the full Hamiltonian of the CmB system is given as follows,
\begin{eqnarray}
\hat{H}&=&\hat{H}_A+\hat{H}_B+\hat{H}_E+\hat{H}_A^{\text{dr}}+\hat{H}_{AB}+\hat{H}_{AE}\cr\cr
&=&\hat{H}_{\bar{A}}+\hat{H}_B+\hat{H}_E+\hat{H}_{AB}+\hat{H}_{AE}
\end{eqnarray}
where the terms are presented in Eqs. (\ref{eq_HA})-(\ref{eq_HAE}) of the main text.

We have combined the $\hat{H}_A$ and $\hat{H}_A^{\text{dr}}$ as the local Hamiltonian of the charger $\hat{H}_{\bar{A}}$. By substituting the $\hat{H}_A$ in Eq. (\ref{eq_HA}) and the $\hat{H}_A^{\text{dr}}$ in Eq. (\ref{eq_Hdr}), it can be rewritten as follows,
\begin{equation}
\hat{H}_{\bar{A}} = \frac{\omega_L}{2}\hat{\sigma}_A^z+\frac{\Delta}{2}\hat{\sigma}_A^z+\frac{\Omega}{2}(\hat{\sigma}^+_Ae^{-i\omega_Lt}+\text{h.c.}),
\end{equation}
where $\Delta=\omega_A-\omega_L$.
In the rotating frame of $\frac{\omega_L}{2}\hat{\sigma}_A^z$, an operator is transformed as $\hat{O}\rightarrow e^{i\frac{\omega_L}{2}\hat{\sigma}_A^zt}\hat{O}e^{-i\frac{\omega_L}{2}\hat{\sigma}_A^zt}$, thus the local Hamiltonian of the charger yields
\begin{equation}
\hat{H}_{\bar{A}}=\frac{\Delta}{2}\hat{\sigma}_A^z+\frac{\Omega}{2}\hat{\sigma}_A^x,
\label{eq_A_HA}
\end{equation}
and the interaction Hamiltonians yield
\begin{equation}
\hat{H}_{AB}=ge^{i\omega_Lt}\hat{\sigma}_A^+\hat{\sigma}_B^-+\text{h.c.},
\label{eq_A_HABs}
\end{equation}
and
\begin{equation}
\hat{H}_{AE}=\sum_{k}{(g_ke^{i\omega_Lt}\hat{\sigma}_A^+\hat{a}_k+\text{h.c.})}.
\label{eq_A_HAEs}
\end{equation}
The free Hamiltonians $\hat{H}_B$ and $\hat{H}_E$ remain invariant.

The eigenvalues of the local Hamiltonian of the charger can be obtained as $\pm\frac{\omega}{2}$ where $\omega=\sqrt{\Delta^2+\Omega^2}$. The eigenbasis are
\begin{equation}
|+_A\rangle=\cos{\frac{\theta}{2}}|\uparrow^z_A\rangle+\sin{\frac{\theta}{2}}|\downarrow^z_A\rangle,
\end{equation}
and
\begin{equation}
|-_A\rangle=\sin{\frac{\theta}{2}}|\uparrow^z_A\rangle-\cos{\frac{\theta}{2}}|\downarrow^z_A\rangle,
\end{equation}
where $\cos{\frac{\theta}{2}}=\frac{\Delta}{\sqrt{\Delta^2+\Omega^2}}$ and $\sin{\frac{\theta}{2}}=\frac{\Omega}{\sqrt{\Delta^2+\Omega^2}}$.

In representation of the eigenbasis $\{|+_A\rangle,|-_A\rangle\}$, the local Hamiltonian of the charger in Eq. (\ref{eq_A_HA}) becomes
\begin{equation}
\bar{H}_{A}=\frac{\omega}{2}\bar{\sigma}_A^z,
\label{eq_A_HA_eig}
\end{equation}
and the raising operator of the charger can be expressed as
\begin{equation}
\hat{\sigma}_A^+=\frac{1}{2}\left(\sin\theta\bar{\sigma}_A^z-\cos{\theta}\bar{\sigma}_A^x-i\bar{\sigma}_A^y\right).
\label{eq_A_spA}
\end{equation}
Note that the operator with a overline means that it lives in the representation of the eigenbasis of $A$.

Now we transform into the interaction picture. Define $\hat{U}_0=\exp{[-i(\hat{H}_{\bar{A}}+\hat{H}_B+\hat{H}_E)t]}$, $\hat{U}=\exp{(-i\hat{H}t)}$ and  $\hat{U}_{I}=\hat{U}^\dagger_{0}\hat{U}$. The density matrix in the interaction picture yields $\rho_I(t)=\hat{U}_I\rho(t_0)\hat{U}^\dagger_I$ and the interaction Hamiltonians in the interaction picture yield respectively,
\begin{equation}
\hat{H}_{AB}^{\text{int}}=\hat{U}_0^\dagger\hat{H}_{AB}\hat{U}_0,
\label{eq_A_HABi}
\end{equation}
and
\begin{equation}
\hat{H}_{AE}^{\text{int}}=\hat{U}_0^\dagger\hat{H}_{AE}\hat{U}_0.
\label{eq_A_HAEi}
\end{equation}

Note that the operators of the charger in Eqs. (\ref{eq_A_HABs}), (\ref{eq_A_HAEs}), (\ref{eq_A_HABi}) and (\ref{eq_A_HAEi})
are represented in the computational basis. In order to derive the master equation of the CmB system, it is convenient to represent the operators of the charger with its eigenbasis by virtue of Eq. (\ref{eq_A_spA}). Consequently, we denote the interaction Hamiltonians in Eqs. (\ref{eq_A_HABi}) and (\ref{eq_A_HAEi}), but with the eigenbasis of the charger, by $\bar{H}_{AB}^{\text{int}}$ and $\bar{H}_{AE}^{\text{int}}$ and the sum by $\bar{H}=\bar{H}_{AB}^{\text{int}}+\bar{H}_{AE}^{\text{int}}$.

In the weak-coupling limit between the charger and the reservoir, one can use the time-convolutionless projection operator technique and obtain the Liouville equation in the interaction picture as follows,
\begin{eqnarray}
\dot{\rho}_I(t)&=&-i[\bar{H},\rho_I(t)]\cr\cr
&=&-i[\bar{H}^{\text{int}}_{AB},\rho(t)]\cr\cr
&&-\int_0^t{[\bar{H}^{\text{int}}_{AE},[\bar{H}^{\text{int}}_{AE}(s),\rho_I(t)]]ds}.
\label{eq_LvNeq}
\end{eqnarray}

The master equation for the CmB system can be derived by tracing over $E$ of Eq. (\ref{eq_LvNeq}). In the limit of a continuum of the reservoir modes $\sum_k{g_k^2}\rightarrow\int{J(\omega)d\omega}$ with the spectral density $J(\omega)$. Assume the spectral density is the Lorentzian as shown in Eq. (\ref{eq_Lorentzian}) and back to the Schr{\"o}dinger picture, the master equation can be obtained as
\begin{equation}
\dot{\rho}_{AB}=-i[\bar{H}+\hat{H}_B+\bar{H}_{AB},\rho_{AB}(t)]+{\cal D}_{A}[\rho_{AB}(t)],
\end{equation}
where the local dissipator ${\cal D}_A$ can be divided into the secular part ${\cal D}$ and the nonsecular part ${\cal D}'$ \cite{haikka2010physcri}. The nonsecular part is given by
\begin{eqnarray}
{\cal D}'[\rho_{AB}]&=&\left[\frac{\gamma_-(t)}{2}-i\kappa_-(t)\right]\{C_-C_0[\bar{\sigma}^+\rho_{AB}(t)\bar{\sigma}^z-\bar{\sigma}^z\bar{\sigma}^+\rho_{AB}(t)]\cr\cr
&&+C_+C_-[\bar{\sigma}^+\rho_{AB}(t)\bar{\sigma}^+-\bar{\sigma}^+\bar{\sigma}^+\rho_{AB}(t)]\}\cr\cr
&&+\left[\frac{\gamma_+(t)}{2}-i\kappa_+(t)\right]\{C_+C_0[\bar{\sigma}^-\rho_{AB}(t)\bar{\sigma}^z-\bar{\sigma}^z\bar{\sigma}^-\rho_{AB}(t)]\cr\cr
&&+C_+C_-[\bar{\sigma}^-\rho_{AB}(t)\bar{\sigma}^--\bar{\sigma}^-\bar{\sigma}^-\rho_{AB}(t)]\}\cr\cr
&&+\left[\frac{\gamma_0(t)}{2}-i\kappa_0(t)\right]\{C_-C_0[\bar{\sigma}^z\rho_{AB}(t)\bar{\sigma}^z-\bar{\sigma}^-\bar{\sigma}^z\rho_{AB}(t)]\cr\cr
&&+C_+C_-[\bar{\sigma}^z\rho_{AB}(t)\bar{\sigma}^+-\bar{\sigma}^+\bar{\sigma}^z\rho_{AB}(t)]\}+\text{h.c.}.
\label{eq_A_nonsecular}
\end{eqnarray}
In Eq. (\ref{eq_parameter_p}) of the main text, the parameter $p$ is defined as the ratio between the correlation time of the reservoir $\tau_C$ and the characteristic time of the charger $\tau_A$. When $p\gg 1$, the nonsecular parts oscillate rapidly with respect to the eigen frequency, i.e. $\omega\sim\tau_A^{-1}$, of the charger and thus can be neglected.

The secular part is given by
\begin{eqnarray}
{\cal D}[\rho_{AB}]&=&C_+^2\gamma_+(t)[\bar{\sigma}_A^-\rho_{AB}(t)\bar{\sigma}_A^+-\frac{1}{2}\{\bar{\sigma}_A^+\bar{\sigma}_A^-,\rho_{AB}(t)\}]\cr\cr
&&+C_-^2\gamma_-(t)[\bar{\sigma}_A^+\rho_{AB}(t)\bar{\sigma}_A^--\frac{1}{2}\{\bar{\sigma}_A^-\bar{\sigma}_A^+,\rho_{AB}(t)\}]\cr\cr
&&+C_0^2\gamma_0(t)[\bar{\sigma}_A^z\rho_{AB}(t)\bar{\sigma}_A^z-\rho_{AB}(t)].
\label{eq_secularD}
\end{eqnarray}
The decay rates $\gamma_\xi$ and $\kappa_\xi$, with $\xi\in\{+,-,0\}$, are determined by the real and imaginary parts of the reservoir correlation function. For the reservoir with a Lorentzian spectral density, the explicit expression of the time-dependent decay rates are obtained in Ref. \cite{haikka2010}. We borrow the results and show them in below
\begin{equation}
\gamma_\xi(t)=\frac{\eta^2}{2(1+q_{\xi}^2)}\left[1-e^{-\lambda t}\cos{(q_{\xi}\lambda t)}+e^{-\lambda t}q_{\xi}\sin{(q_{\xi}\lambda t)}\right],
\end{equation}
\begin{equation}
\kappa_\xi(t)=\frac{\eta^2}{1+q_\xi^2}\left[-q_{\xi}+e^{-\lambda t}q_{\xi}\cos{(q_{\xi}\lambda t)}+e^{-\lambda t}\sin{(q_{\xi}\lambda t)}\right],
\end{equation}
where $q_{\xi}=s-\xi p$. The behavior of the decay rate for $p=100$ and $s=6$ is shown in Fig. \ref{fig2_decayrate}(a) of the main text.  One can see that the amplitude of $\gamma_{\pm}$ is so small compared with $\gamma_0$ that can be neglected.

For the resonant case $\Delta=0$, the eigenbasis is specialized to $|+_A\rangle=|\uparrow^x_A\rangle$ and $|-_A\rangle=|\downarrow^x_A\rangle$ and the coefficient are given by $C_+=C_0=\frac{1}{2}$ and $C_-=-\frac{1}{2}$. The local Hamiltonian in Eq. (\ref{eq_A_HA}) reduces to $\hat{H}_{\bar{A}}=\frac{\Omega}{2}\hat{\sigma}_A^x$, while the eigen frequency is simplified as $\omega=\Omega$, indicating that $\bar{\sigma}^z_A=\hat{\sigma}_A^x$. Therefore, the dissipator in Eq. (\ref{eq_secularD}) reduces to Eq. (\ref{eq_dissipator}) in the main text.

\section{The charging time and the maximum ergotropy}
\label{sec_app_B}
\begin{figure}[!ht]
  \includegraphics[width=.95\linewidth]{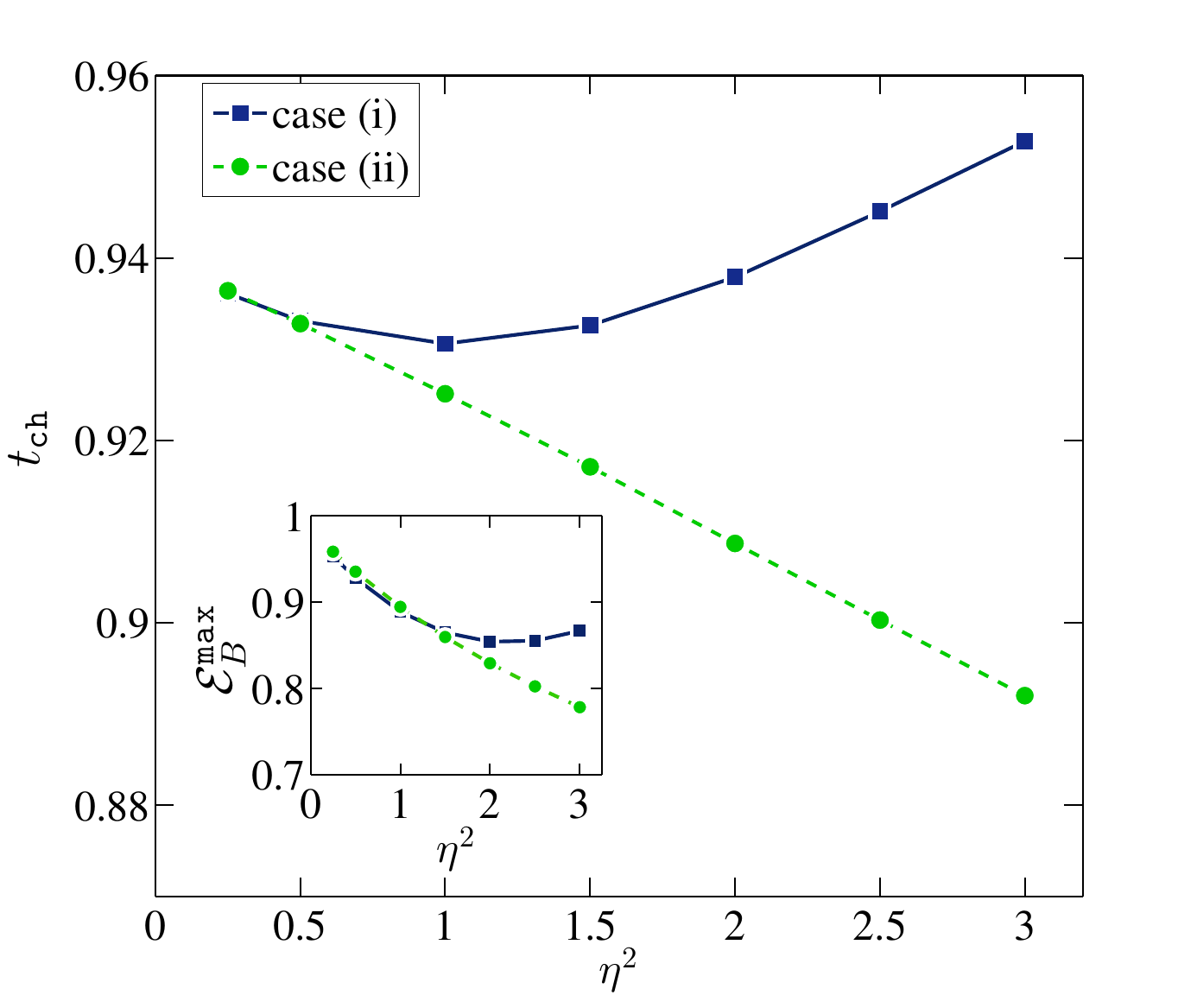}
  \caption{The charging time versus the coupling strength for cases (i) and (ii). The inset shows trends of the maximum ergotropy with the coupling strength. The parameters are chosen to be the same as in Fig. \ref{fig3_ergotropy}.}
  \label{fig6_chargingtime}
\end{figure}

In this section, we provide a supplementary explanation regarding the results shown in Fig. \ref{fig3_ergotropy}. As mentioned in the main text, the charging time $t_{\text{ch}}$ refers to the time at which the ergotropy reaches its maximum ${\cal E}_B^{\text{max}}$. In Fig. \ref{fig6_chargingtime}, we show the numerical results of the charging time with various coupling strength, characterized by $\eta^2$ in Eq. (\ref{eq_gamma_t}), between the charger and the reservoir, for cases (i) and (ii). One can see that the $t_{\text{ch}}$ is shorten as $\eta$ increases in case (ii). Meanwhile, as shown by the circles in the inset of Fig. \ref{fig6_chargingtime}, the ${\cal E}_B^{\text{max}}$ of case (ii) is also lowered, indicating that the shortening of the charging time is at the cost of the maximum ergotropy. The $t_{\text{ch}}$ of case (i) varies nonmonotonically as $\eta$ increases. Overall, for identical coupling strength $\eta$, the $t_{\text{ch}}$ of case (i) is always slightly longer than that of case (ii).

\end{appendix}

\end{document}